\documentclass{article}
\usepackage{graphicx}
\usepackage{graphics}
\title{The Effects of Spin-Excitons on the Surface States of SmB$_6$: A Photoemission Study.}

\author{Arian Arab${^{1,2}}$, A.X. Gray${^{1,2}}$,\\
S. Nemsak$^3$, D.V. Evtushinsky$^{4,5}$,\\ C.M. Schneider$^{3}$,\\
Dae-Jeong Kim$^6$, Zachary Fisk${^{6}}$,\\
P.F.S. Rosa${^{7}}$, T. Durakiewicz${^{7,8}}$\\ and\\ P.S. Riseborough${^{1,2}}$\\
${^1}$Physics Department, Temple University,\\ Philadelphia, Pa 19122\\
${^2}$Temple Materials Institute, Temple University,\\ Philadelphia, Pa 19122\\
$^3$Peter-Grunberg-Institut 6, Forschungszentrum Julich GmbH\\ 52425 Julich, Germany\\
$^4$Helmholtz-Zentrum Berlin fur Materialien und Energie GmbH,\\ Berlin, Germany\\
$^5$IFW-Dresden, Helmholtzstrasse 20,\\ 01069 Dresden, Germany\\
${^6}$Department of Physics and Astronomy,\\ University of California, Irvine, Ca 92697\\
${^{7}}$Los Alamos National Laboratory,\\ Los Alamos, NM 87545\\
${^8}$Institute of Physics, Maria Curie-Sklodowska University,\\ 20-031 Lublin, Poland}

\begin{document}

\maketitle

\begin{abstract}
We present the results of high-resolution valence-band photoemission spectroscopic study of SmB$_6$ which shows evidence for a V-shaped density of states of surface origin within the bulk gap. The spectroscopy data is interpreted in terms of the existence of heavy 4$f$ surface states, which may be useful in resolving the controversy concerning the disparate surface Fermi-surface velocities observed in experiments. Most importantly, we find that the temperature dependence of the valence-band spectrum indicates that a small feature appears at a binding energy of about - 9 meV at low temperatures. We attribute this feature to a resonance caused by the spin-exciton scattering in SmB$_6$ which destroys the protection of surface states due to time-reversal invariance and spin-momentum locking. The existence of a low-energy spin-exciton may be responsible for the scattering which suppresses the formation of coherent surface quasi-particles and the appearance of the saturation of the resistivity to temperatures much lower than the coherence temperature associated with the opening of the bulk gap.
\end{abstract}

\section{Introduction}

The material SmB$_6$ is a Kondo insulator \cite{Menth} in which a hybridization gap in the bulk density of states of about 21 meV forms for temperatures below a coherence temperature of about 100 - 150 K \cite{Miyazaki,Zhang}. The small magnitude of the gap is due to a renormalization by the strong Coulomb interactions between the Sm $4f$ electrons. Bulk impurity states pin the chemical potential in the gap. Thus, the bulk properties are characterized by a thermal activation in the resistivity and specific heat with an activation energy of about 4 meV \cite{Flachbart,Gorshunov,Zhou}. However, at temperatures of the order of $4$ K the resistivity shows a plateau \cite{Nickerson,Allen} indicative of the presence of surface states at the Fermi-energy \cite{Wolgast,Kim1,Zhang,Kim2,Syers}. The material SmB$_6$ has been the focus of much renewed interest ever since it was proposed that SmB$_6$ is a topological Kondo Insulator \cite{Dzero,Takimoto}. Doping SmB$_6$ with magnetic impurities has been shown to increase the low temperature resistivity, whereas non-magnetic impurities do not change the resistivity saturation \cite{Kim2}. The insensitivity to non-magnetic impurities suggests that the electronic surface states are protected from $\underline{k} \rightarrow -\underline{k}$ scattering by a spin texture and time-reversal symmetry as expected for a topological insulator. Furthermore, it has been reported \cite{Xu2} that a spin texture has been observed in the above Fermi-surface states by a spin-resolved ARPES experiment. The reported spin texture is expected for a topological insulator \cite{Schlottmann}. However, the spin-structure of the surface states is disputed \cite{Hlawenka}.

The existence of the metallic surface states found in transport measurements  is also consistent with the results of magnetic torque \cite{Li} and ARPES experiments \cite{Jiang,Neupane,Xu}, although alternate three-dimensional interpretations have been proposed \cite{Tan,Frantzekas}. The measured surface Fermi-surface consists of three sheets \cite{Jiang,Neupane,Xu}; a small sheet around the $\Gamma$-point, and elliptical sheets located around the $X$ and $Y$-points. ARPES measurements \cite{Neupane} have identified an in-gap peak due to surface states at an energy of about - 5 meV. Surface Fermi-surface velocities as large as $v_F \approx 8.45 \times 10^5$ m/s have been observed in Quantum Oscillation measurements while ARPES measurements yield $v_F \approx 4.0 \times 10^4$ m/s which are an order of magnitude lower. Both of these values are orders of magnitudes larger than those obtained from theory \cite{Alexandrov,Lu} ($v_F \sim 4.5 \times 10^3$ m/s, $v_F \sim 7.6 \times 10^3$ m/s) and also from transport measurements which yield $v_F \approx 7 \times 10^2 $ m/s \cite{Luo}. The large discrepancies between the values of the surface Fermi-velocities have led to the speculation that the surface may have undergone a phase transition in which the 4f ions localize leaving the surface states to be dominated by conduction electrons with light masses \cite{Alexandrov&Onur}.

Here we present the results of high-resolution angle-integrated photoemission spectroscopy measurements of SmB$_6$ which indicate the presence of a V-shaped density of states within the bulk gap. We estimate that the Weyl-point and the chemical potential reside within the bulk gap and have very similar energies. This points to the existence of a heavy band of surface electronic states, as has been predicted by theory and inferred from low-temperature transport measurements \cite{Luo}. By examining the difference of the integrated spectra at $T=20$ K and $T=1.2$ K, we find that the density of states exhibits a low temperature peak located at about - 9 meV. The difference spectra shows that this feature has its intensity derived from the $X$ and $Y$-points, and is of surface origin. We identify this peak as a resonance in the surface electronic density of states caused by the coupling of the surface states to the in-gap spin-exciton excitations \cite{Kapilevich}.Spin-excitons are magnetic excitations that occur in the gap of paramagnetic Kondo Insulators \cite{Riseborough1,Riseborough2,Riseborough3}, and have been observed in bulk SmB$_6$ with excitation energies of 12 to 14 meV \cite{Alekseev1,Alekseev2,Alekseev3,Fuhrman}. The intensity of the spin-exciton peak seen in neutron scattering experiments \cite{Alekseev3} increases and the peak width decreases with decreasing temperature. At the temperature of 25 K the intensity is approximately half of its value at liquid Helium temperatures while its line width is saturating at a value comparable to the experimental resolution \cite{Alekseev3}. Supporting evidence for the exceptionally long lifetime of the spin-exciton has been provided by  recent Raman scattering experiments \cite{Valentine} which see a $q \sim 0$ feature at 16 meV with a width of 0.5 meV in an Al flux-grown SmB$_6$ sample at $T  =  15 $ K. The narrow line-width of the spin-exciton is caused by the absence of an electron-hole pair decay channel within the bulk hybridization gap. A shift of the peak from 13 meV in the bulk to 9 meV at the surface would indicate that the surface states are close to a quantum critical point \cite{Riseborough4}. The existence of large-amplitude, low-frequency spin-flip scattering at the surface would also result in the surface states not being completely protected from back-scattering and gives rise to a resonant peak in the low-temperature surface electronic density of states. This spin-exciton induced resonance in the surface electronic spectrum is expected to occur at low-temperatures at a fairly sharply defined energy but, due to the requirement of conservation of crystal momentum combined with the fairly narrow spin-exciton dispersion, should be spread over a finite range of crystal momenta. Since the resonance is a many-body effect, it is also expected to have a significantly reduced spectral weight. Therefore, this feature is expected to be most identifiable as the difference between the high temperature and low temperature angle-integrated photoemission spectrum.

\section{Experiment and Results}

High-quality SmB6 crystals were grown using aluminum flux method in a continuous Ar purged vertical high temperature tube furnace [9]. High-resolution valence-band photoemission measurements were carried out in the normal emission geometry using linearly p-polarized light with photon energy of 35 eV at the $1^3$ end-station of the BESSY II storage ring of the Helmholtz-Zentrum Berlin. The sample was cleaved in-situ at 1.2 K along the (001) plane and measured in an ultra-high vacuum of 10$^{-11}$ torr at cryogenic temperatures of 1.2 K and 20 K. The overall energy resolution was estimated to be about 3 meV.

\begin{figure}[!ht]
  \centering
  \includegraphics[width=10cm]{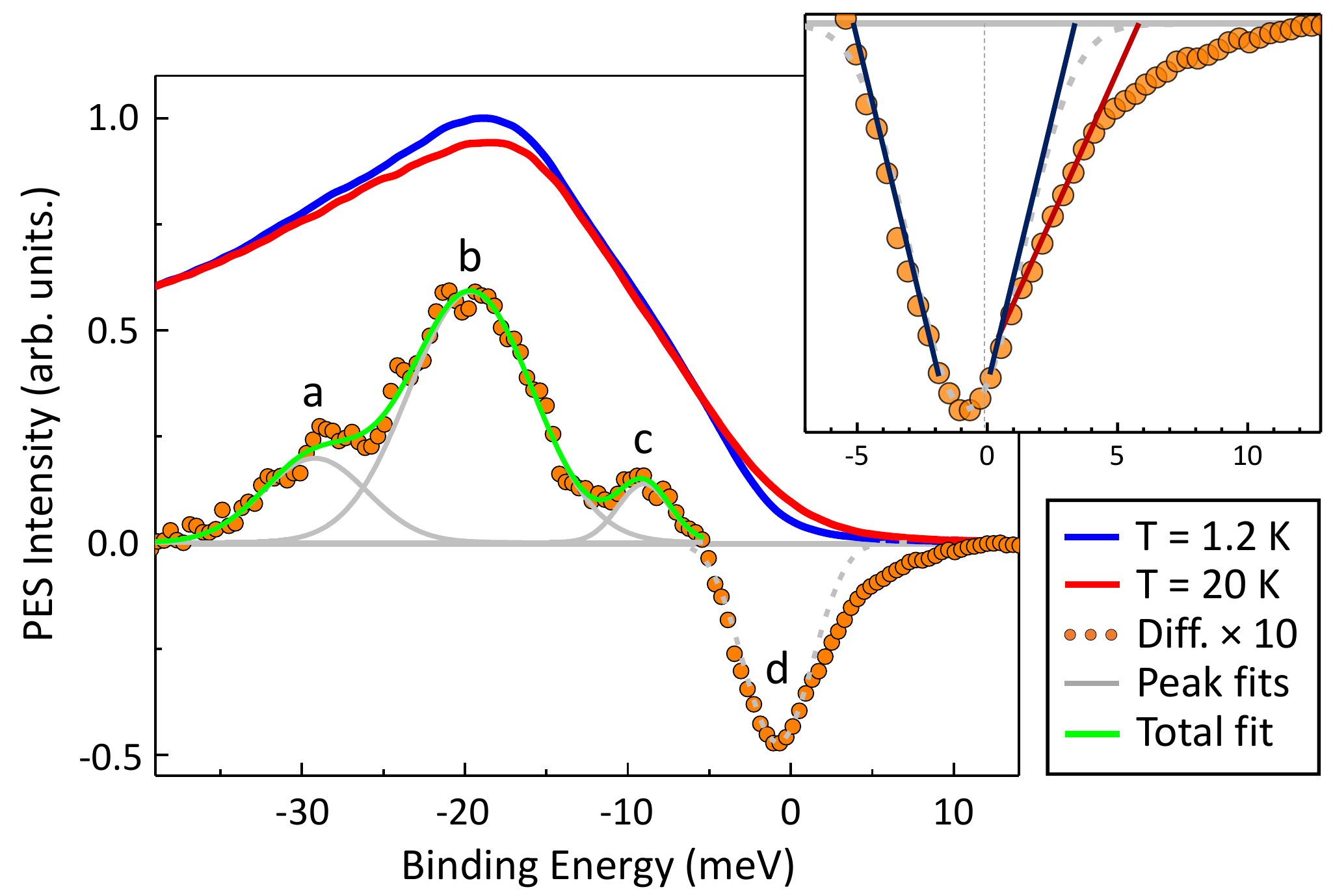}
  \caption{The angle integrated valence-band photoemission spectra measured at the sample temperatures of 1.2 K (blue line) and 20 K (red line).  The difference between the spectra at these two temperatures is shown in orange. To enhance the excursions, the difference spectrum has been multiplied by a factor of 10. The inset shows a close up of the difference spectrum near the Fermi-energy. Spectra were measured along the $\overline{X}-\overline{\Gamma}-\overline{X}$ direction in the BZ and integrated over the entire angular range of the detector (30 deg). (Colour online)}
\label{Fig1}
\end{figure}

Fig(\ref{Fig1}) shows the intensity profiles of the angle-integrated valence-band photoemission spectra measured at the sample temperatures of 1.2 K (blue) and 20 K (red).  The difference spectrum is shown in orange. To enhance the major features, the spectrum has been multiplied by a factor of 10.  The three positive peaks below - 5 meV, labeled as (a), (b) and (c), were fit with simple Gaussians. Feature (d), near the Fermi-energy has a negative weight and cannot be fit with a single Gaussian lineshape.\\

Feature (a) occurs at the binding energy of - 29 meV. The presence of a low-temperature surface feature at this energy is completely unexpected. We note that this feature is separated from the edge of the bulk gap by an energy that roughly corresponds to the spin-exciton energy. This might indicate that the feature is a bulk feature caused by coupling to a spin-exciton, however such a temperature dependence is not expected for a bulk feature. Alternatively, this feature might be attributable to a crystal field level. However, a corresponding crystal field feature was not observed in inelastic neutron scattering experiments and would not be expected to be so temperature dependent.\\

Feature (b) is a non-dispersive peak located at a binding energy of about 20 meV. It is identified as a 4$f$ feature at the bottom of the bulk gap caused by the hybridization of Sm 4$f$ and Sm 5$d$ electronic states. The bulk ARPES of the Kondo insulator SmB$_6$ has been previously reported \cite{Miyazaki} and the bulk gap has been observed to form below a characteristic temperature of between 80 K and 150 K. \\

Feature (c)  at - 9 meV is present at the lower temperature (1.2 K) but is absent at the higher temperature (20 K). This feature is of surface origin and its weight is mainly derived from the vicinity of the $X$ and $Y$ points [23]. This feature occurs at the a similar energy as the - 8 meV feature observed by Miyazaki {\it et al.} \cite{Miyazaki}. Due to its energy, its temperature dependence and its appearance in the angle integrated spectrum, we attribute this feature to a resonance in the surface electronic density of states caused by the coupling of the surface states to spin-exciton excitations \cite{Kapilevich}. More specifically, it is assigned to a resonant process involving the filling of the photo-hole at $\omega=-\omega_0$ by an electron in a state just below $\omega=0$ accompanied by the emission of a spin-exciton of energy $\omega_0$. Because the process involves electrons just below the Fermi-energy, the resonant feature depends sensitively on temperature. \\

Feature (d) centered at - 0.8 meV resides within the bulk gap and has a V-shape. It can be identified with the conical surface density of states with a minimum that is located at the Weyl-point  \cite{Greene}. The peak has been fit (gray dashed-line) with the convolution of a V-shaped density of states and a Gaussian resolution function of the full-width half-maximum of 3 meV (the overall experimental resolution). We note that the shape of feature (d), shown in the inset of Fig(\ref{Fig1}), resembles the differential conductance reported by Park {\it et al.}  \cite{Greene}. The data of Park {\it et al.} is of the form of an asymmetric V-shape followed by a break a distinct flattening at + 4 meV, which is related to a surface density of states coming from the inequivalent cones located at the $\Gamma$ and $X$-points and to the coupling to a spin-exciton with energy of 4 meV. The straight lines are guides to the eye.\\

The surface electronic density of states can be estimated by subtracting a Shirley background from the angle-integrated spectrum and then normalizing the area below the Weyl point to unity. From the slope of the normalized data, we estimate that the Fermi-velocity at the temperature of 1.2 K is about $5.2 \times10^3$ m/s, while at the higher temperature of 20 K the Fermi-velocity is estimated to be $5.8 \times 10^3$ m/s. These values are compatible with the theoretical values \cite{Alexandrov,Luo} of $v_F = 4.5 \times 10^3$ m/s and $v_F = 7.6 \times 10^3$ m/s.

Reference \cite{Neupane} reported features appearing similar to those found here, and residing around -5 meV within a gap of 15 meV. An in-gap dispersive-band was also reported in Reference \cite{Xu3}. The authors note that the features discussed in references \cite{Neupane} and \cite{Xu3} were interpreted as having a different origins from the interpretation discussed here, but the features are somewhat similar to feature (c) in our angle-integrated spectrum, notably with in-gap features developing at or below 20K. We also note that the in-gap peaks are not well resolved. Here, we identify the in-gap peak as a resonance in the surface density of states caused by spin-exciton scattering \cite{Kapilevich}, and its intensity originates from a broad region around the $X$ and $Y$ points, since the Weyl cones are centered at these points. Like feature (d), the ARPES spectrum found by Xu {\it et al.} at $k_F$ also shows an almost linear decrease in intensity as a binding-energy of $\sim - 2$ meV is approached. Since the decrease occurs over an energy scale of 5 meV, even at the lowest temperature of 1.2 K, we believe that the observed decrease is not due to the effects of the Fermi-function but instead is due to the presence of an electronic scattering process and the minimum of the spectral density associated with the Weyl-point, as inferred from feature (d) in our difference spectrum.

\section{Theory}

In order to make closer contact with experiment we consider the (unperturbed) Weyl cone to be asymmetric, where the Weyl point is separated from the upper edge of the hybridization gap by an energy, $\Delta_+$, of only + 5 meV and from the lower edge by an energy $\Delta_-$ of 18 meV. Since we are interested in the evolution of the surface electronic states for temperatures below 25 K, it seems reasonable to assume that the temperature dependency of the hybridization gap has saturated and, thus, $\Delta_{\pm}$ may be approximated by constant values.  The values of $\Delta_{\pm}$ were chosen so as to model experimental data. The surface Fermi-energy is positioned at an energy $\mu =  +  2$ meV above the Weyl point. This positioning of the surface Fermi-energy is consistent with the $\approx $ 3 meV activation energy observed in bulk experiments \cite{Flachbart} which is associated with bulk impurities that pin the Fermi-energy. The existence of these bulk in-gap states has recently been confirmed by Raman scattering measurements \cite{Valentine} which indicate that 1 \% of Sm vacancies close the gap in the bulk density of states. This vacancy concentration is in very good agreement with theoretical predictions \cite{Schlottmann1,Schlottmann2} that suggests that the bulk hybridization gap is closed with only 5 \% of Sm vacancies \cite{Riseborough3}. Since the inelastic neutron scattering experiments indicate that the spin-exciton energy $\hbar \omega(\underline{q})$ has only a weak dispersion \cite{Alekseev1,Alekseev2,Alekseev3,Fuhrman}, we shall mainly ignore the dispersion. We shall set the spin-exciton energy as $\hbar \omega_0 \ \sim 8 $ meV, which is reduced from the bulk value since, due to reduced coordination number, the surface is expected to be closer to a magnetic instability.

\begin{figure}[!ht]
  \centering
  \includegraphics[width=5cm]{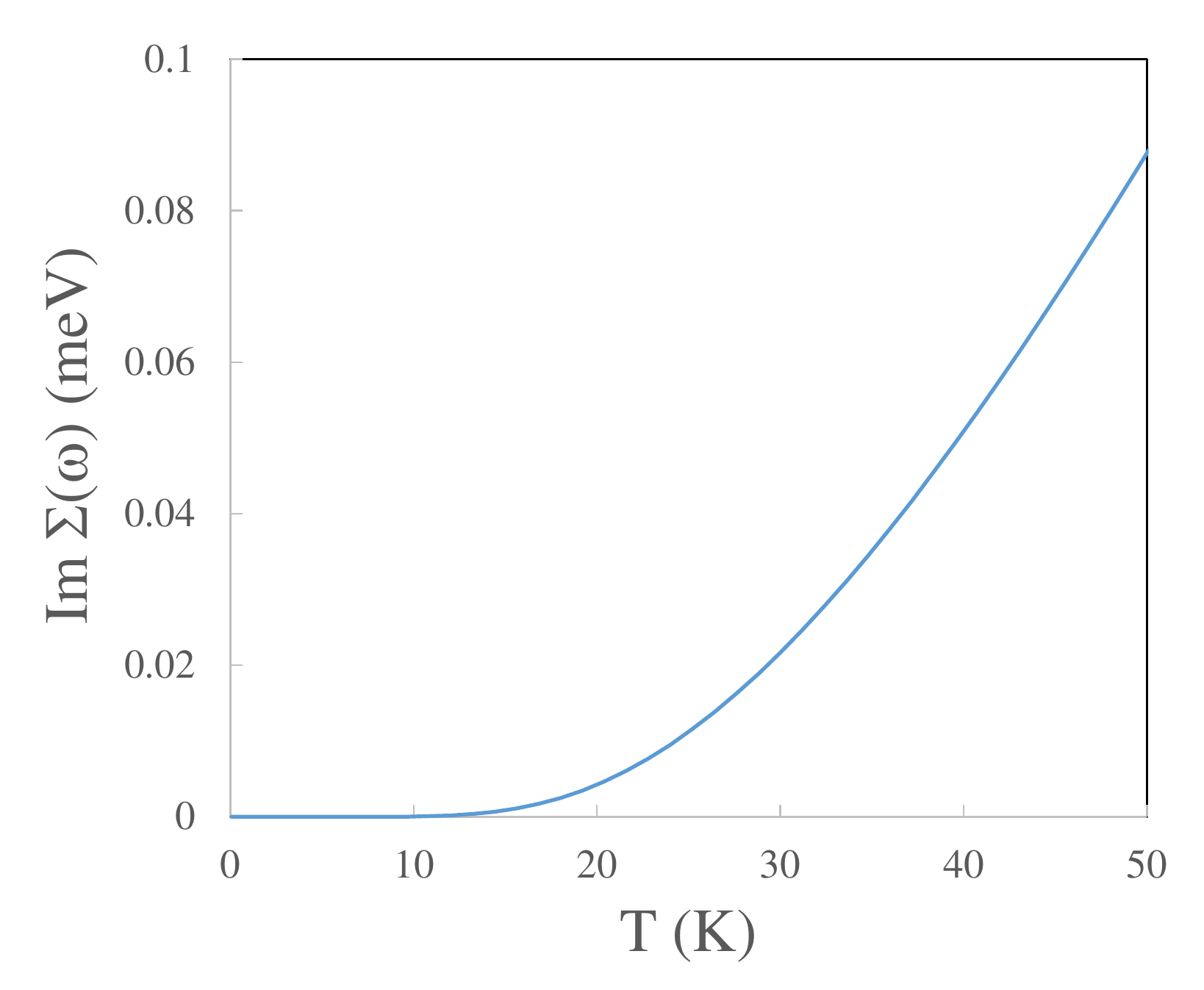}
  \caption{The thermally activated temperature dependence of the imaginary part of the self energy due to interactions with spin-excitons, for electrons on the surface Fermi-surface.  (Colour online)}
\label{Fig2}
\end{figure}

We calculate the self energy for electrons in the surface states due to the emission and absorption of spin-excitons \cite{Kapilevich}, using the imaginary time Green's functions and then analytically continuing $i \ \omega_n \ \rightarrow \ \omega - \ i \ \eta$. At $T=0$, the self energy only has a contribution from the fermionic statistics and is evaluated as
\begin{eqnarray}
  \Re e \ \Sigma(\omega) & = &  \bigg({3 J'^2 \over 4 Z_0}\bigg) \ \bigg[ \ \bigg({ \Delta_+ - \Delta_- \over \Delta_+ \Delta_-} \bigg) + {x_- \over \Delta_+^2} \ln \bigg{\vert} \ {\hbar (\omega - \omega_0) \over x_- -  \Delta_+ } \bigg{\vert}  \nonumber \\
  && \ - \ {x_+ \over \Delta_+^2} \ln \bigg{\vert} \ {\hbar (\omega + \omega_0) \over x_+   } \bigg{\vert} + {x_+ \over \Delta_-^2} \ln \bigg{\vert} { x_+ \over x_+ + \Delta_-} \bigg{\vert} \ \bigg] \nonumber \\
  \Im m \ \Sigma(\omega) & = & \bigg({3 \pi J'^2 \over 4 Z_0}\bigg) \ \bigg[  + {x_- \over \Delta_+^2}  \ \bigg{\{} \Theta(\Delta_+ - x_-) -  \Theta (\omega_0-\omega) \bigg{\}} \nonumber \\
  && \ + \ {x_+ \over \Delta_+^2} \  \bigg{\{} \Theta(x_+) -  \Theta (\omega_0+\omega) \bigg{\}} \ - \ {x_+ \over \Delta_-^2}  \ \bigg{\{} \Theta(\Delta_- + x_+) -  \Theta (x_+) \bigg{\}}  \ \bigg] \nonumber \\
\end{eqnarray}
where
\begin{equation}
  x_{\pm}  =  \hbar ( \omega \pm \omega_0)  +  \mu
\end{equation}
and where $Z_0^{-1}$ is a dimensionless quantity that determines the intensity of the spin-exciton. The weighting factor $Z_0$ is given by
\begin{equation}
  Z_0 = J^2 \bigg( {\partial \chi^0(\omega)\over \partial \hbar \omega}\bigg)\bigg{\vert}_{\omega_0} \ \sim \  {2 J^2 \hbar \omega_0 \over ( \Delta_+ + \Delta_-)^3}
\end{equation}
where $J$ is the value of the bulk exchange interaction. We shall use the ratio of the surface to bulk values of the Kondo exchange of
\begin{equation}
{J'\over J} \  \sim \ \exp\bigg[-{(\Delta_+ + \Delta_-) \over \hbar v_F} \bigg]
\end{equation}
which is estimated as $\sim \exp[-8/3]$.

The real part of the $T=0$ self energy has logarithmic singularities at $\omega \ = \ \pm \ \omega_0$ with coefficients proportional to the unperturbed density of states at the Fermi-energy (i.e. $\sim {2 \mu / \Delta_+^2}$). It is these peaks in the self energy that are responsible for the resonant structure in the angle integrated density of states and, since the structure is associated with electrons or holes in the close vicinity of $\mu$, the intensity of the structure is expected to decrease as the temperature is raised because of smearing caused by the Fermi-function \cite{Kapilevich}. The inclusion of a finite dispersion for the spin-exciton energies would also result in the $T=0$ logarithmic singularities being broadened \cite{Kapilevich}. The logarithmic singularities involving $\Delta_{\pm}$ are artifacts due to the continuum model having discontinuities in the density of states at the band edges. The structures associated with the band edges, in a more realistic model \cite{Kapilevich}, should consist of rounded peaks. The imaginary part of the $T=0$ self energy is non-zero for energies $\omega$ such that  $\omega \ge \omega_0$ or $- \ \omega_0 \ \ge  \omega $, and so it doesn't contribute to scattering of quasi-particles near the Fermi-energy. For finite temperatures, the fermionic contribution becomes appreciable outside the edges of these regions due to Fermi-function smearing. However, the bosonic contribution to the self energy is non-zero at the Fermi-energy $\hbar \omega=0$ and has an intensity proportional to the Bose-Einstein distribution function $N(\hbar \omega_0)$. The bosonic contribution to the self energy, $\Sigma_B(\omega)$, is given by
\begin{eqnarray}
\Re e \ \Sigma_B(\omega) & = & N(\hbar\omega_0) \ \bigg({3 J'^2 \over 4 Z_0}\bigg) \ \bigg[ \ 2 \bigg( { \Delta_+ -\Delta_- \over \Delta_+ \Delta_-} \bigg) - {x_+ \over \Delta_+^2} \ln \bigg{\vert} \ {x_+-\Delta_+ \over x_+ } \bigg{\vert}  \nonumber \\
&& \  - {x_- \over \Delta_+^2} \ln \bigg{\vert} \ {x_- -\Delta_+ \over x_- } \bigg{\vert}   - {x_+ \over \Delta_-^2} \ln \bigg{\vert} \ {\Delta_- + x_+ \over x_+ } \bigg{\vert} \nonumber \\
&& \  - {x_- \over \Delta_-^2} \ln \bigg{\vert} \ {\Delta_- + x_- \over x_- } \bigg{\vert} \ \bigg] \nonumber \\
\Im m \ \Sigma_B(\omega) & = & N(\hbar\omega_0) \ \bigg({3 \pi J'^2 \over 4 Z_0}\bigg) \ \bigg[ \ { x_- \over \Delta_+^2} \bigg{\{} \Theta(\Delta_+ - x_-) -  \Theta (-x_-) \bigg{\}} \nonumber \\
&& \ +  \ { x_+ \over \Delta_+^2} \bigg{\{} \Theta(\Delta_+ - x_+) -  \Theta (-x_+) \bigg{\}} - { x_- \over \Delta_-^2} \bigg{\{} \Theta(\Delta_- + x_-) -  \Theta (x_-) \bigg{\}}  \nonumber \\
&& \ - \ { x_+ \over \Delta_-^2} \bigg{\{} \Theta(\Delta_- + x_+) -  \Theta (x_+) \bigg{\}} \ \bigg] \ .
\end{eqnarray}
As can be seen in Fig(\ref{Fig2}), the imaginary part of the bosonic contribution to the self energy of quasi-particles at the Fermi-energy is thermally activated. For the parameters used in this model, one finds that at $50$ K where the spin-exciton intensity starts to be appreciable and its width is still a decreasing function of temperature \cite{Alekseev3}, the quasi-particle scattering time is estimated to be of the order of $ 9 \times 10^{-12}$ seconds.

However, for temperatures of the order 10 K where the spin-exciton intensity has saturated and its line width is resolution limited, the spin-exciton scattering rate is almost completely suppressed since $ \tau \sim 3 \times 10^{-8}$ seconds. For this temperature, one expects that the dominant contribution to the surface electron scattering comes from interactions with magnetic and non-magnetic impurities. This type of temperature variation of the scattering rate is roughly consistent with recent ultrafast terahertz spectroscopy measurements on SmB$_6$ \cite{Averitt}. The experimentally determined scattering rate, ($1 /2 \tau$), shows a very rapid decrease from a value of $8 \times 10^{12}$ seconds$^{-1}$ at a temperature of about 30 K and a subsequent saturation at a value of roughly $4 \times 10^{12}$ seconds$^{-1}$ at $T \approx 10$ K.
\begin{figure}[!ht]
  \centering
  \includegraphics[width=5cm]{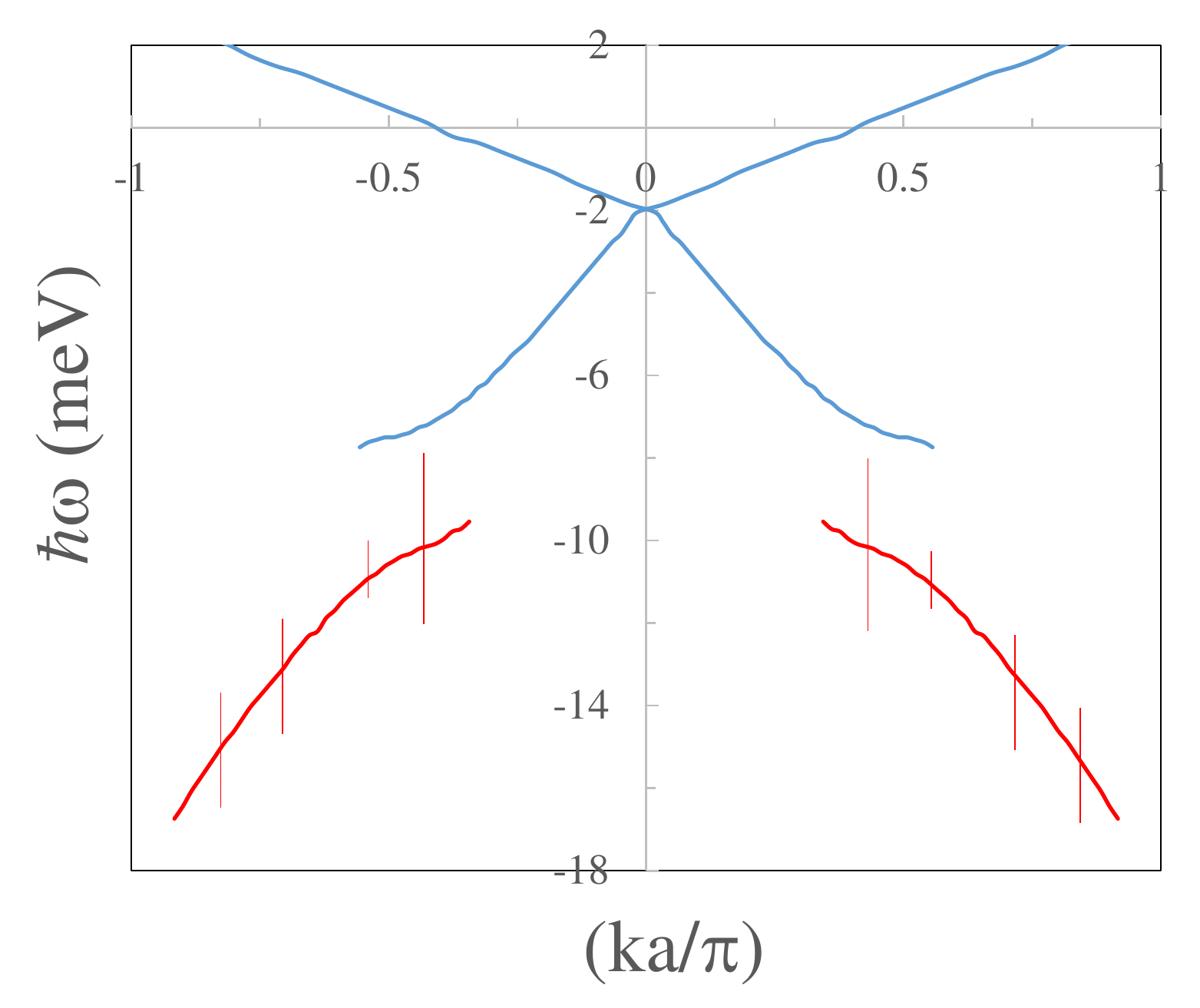}
   \includegraphics[width=5cm]{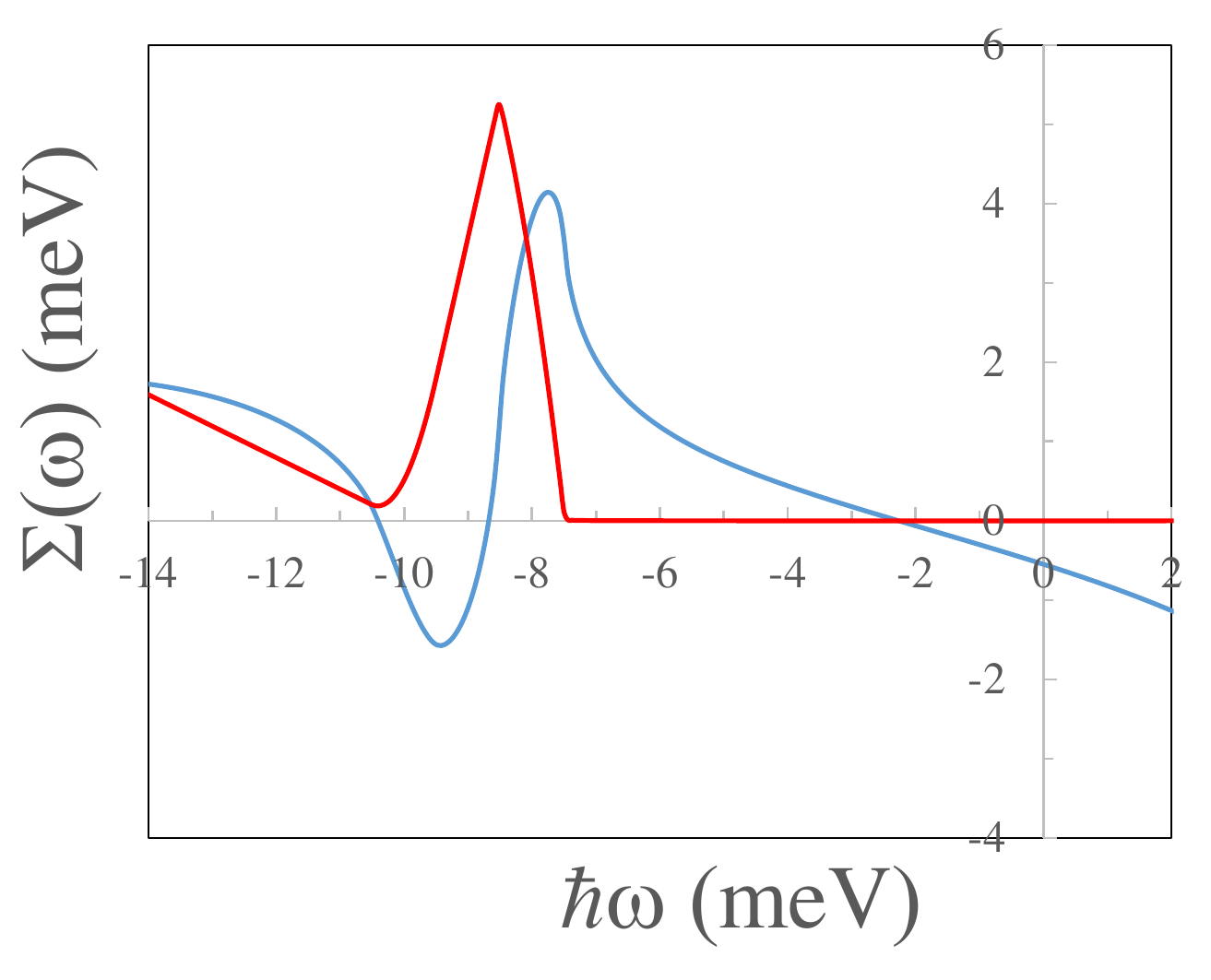}
  \caption{The calculated quasi-particle $T=0$ dispersion relations along the diagonal line $(k,k)$ of the two-dimensional surface Brillouin zone (Left) and the frequency-dependence of the $T=0$ self energy $\Sigma(\omega)$ (Right). The real part of the self energy is shown in blue and the imaginary part is shown in red.  (Colour online)}
\label{Fig3}
\end{figure}

The $T=0$ quasiparticle dispersion relation calculated from
\begin{equation}
  \hbar \ \omega \ = \ \epsilon_{\tau}(\underline{k}) \ - \ \mu \ + \ \Re e \ \Sigma_{\tau}(\underline{k},\omega) \ ,
\end{equation}
in which the spin-exciton is given a dispersion of 1 meV, is shown in the left panel of Fig(\ref{Fig3}). It is seen that the states with chirality index $\tau=-1$ exhibits a strong mass enhancement for energies just above the spin-exciton resonance (blue), where the imaginary part of the self energy is thermally activated. However, although the states below the spin-exciton energy also show a similar mass enhancement (red), these states do not result in simple quasi-particle peaks as the imaginary part of the self-energy (right panel) is large and highly frequency dependent. The interaction with the spin-exciton transfers spectral weight from the near Fermi-energy quasiparticle states to the continuum of incoherent excitations indicated in the left panel by the vertical lines. It is interesting to note that the resonance feature, although removed from the Fermi-energy by an energy $\hbar \omega_0$, occurs for a range of wave vectors spanning $k \sim k_F$. (See panel c of Fig(4) in reference \cite{Xu3}.)

\section{Discussion}

We have observed an in-gap feature in the photoemission spectra at an energy - 9 meV, as did Miyazaki {\it et al.} \cite{Miyazaki}.  However, other ARPES experiments \cite{Neupane,Xu3} observed an in-gap surface feature at around - 5 meV. This discrepancy is compatible with the overall experimental resolution combined with the difficulty of accurately determining the Fermi-energy, but the discrepancy might also be due to the measurements being made on different types of surface areas. This is quite probable since Rossler {\it et al.} have observed \cite{Rossler}, through STM measurements, that reconstructed and unreconstructed patches with different terminations may be found on the same surfaces.

Similar in-gap resonances have been previously observed in STM measurements by Ruan {\it et al.} \cite{Ruan}. The features observed by Ruan {\it et al.} had a similar temperature dependence to that inferred for feature (c). Ruan {\it et al.} attributed the resonance to the effect of collective magnetic fluctuations, similar to the spin-exciton mechanism advocated here. The - 9 meV energy of the feature that we observe is also consistent with the - 8 meV feature seen in the STM measurements of Yee {\it et al.} \cite{Yee}. The peak observed by Yee {\it et al.} is damped with increasing temperature much faster than expected from thermal broadening. In fact, the intensity of the feature was completely suppressed above $T \ \sim 40$ K. The experiment of Yee {\it et al.} also showed a residual spectral density within the gap, in accord with the theory of Kapilevich {\it et al.} \cite{Kapilevich}. However, Yee {\it et al.} argue that the - 8 meV feature is due to the effect of the bulk hybridization. We consider this explanation as unlikely since the bulk 20 meV gap already starts forming in the temperature range of  80 K \cite{Greene} $\sim$ 100 K \cite{Zhang} $\sim$ 150 \cite{Miyazaki}. We consider it more likely that the sharp - 8 meV peak observed by Yee {\it et al.} and the feature reported here have a common origin and represents a spin-exciton resonance in the surface Weyl-cone density of states \cite{Kapilevich}.

Further support for the picture advocated here is given by the measurements by Park {\it et al.} \cite{Greene} of planar tunneling between SmB$_6$ and superconducting Pb. The tunneling measurements showed the growth of surface states in the temperature range $20 \ \sim \ 15 $ K, and a V-shaped density of states expected from a Weyl cone, with only a small 0.2 meV separation of the Weyl point from the Fermi-energy. The close proximity of the putative Weyl point and Fermi-energy together with the experimentally observed $X$-point Fermi-surface $k_F$ value of 0.3 A$^{-1}$ indicate Fermi-surface velocities which are comparable to those inferred from the transport measurements of Luo {\it et al.} \cite{Luo}. Such a heavy surface Fermi-surface sheet should prove extremely difficult to observe in either de Haas - van Alphen or ARPES measurements. The linearity of the density of states observed by Park {\it et al.} ended at $\pm \ 4 $ meV. Furthermore, Park {\it et al.} found an inelastic feature at $- \ 4$ meV which is attributed to inelastic tunneling involving spontaneous emission of spin-excitons. The difference in the Fermi-energies and the positions of the inelastic features may be caused by the different treatment of the surfaces. In particular, the tunneling barriers were formed by plasma oxidization of the polished surfaces which may pacify any dangling bonds \cite{Rossler}. The characteristic energy of - 4 meV is compatible with the energy of the in-gap surface feature found through ARPES by Neupane {\it et al.}  \cite{Neupane} who also found that well-defined surface features only developed below a temperature of about $15$ K, in agreement with the results of Park {\it et al.} \cite{Greene} and the results presented here.

We note that the observation of patches of reconstructed surface implies that different patches have different properties. Since the values of spin-exciton energies apparently range from 14 meV in the bulk and from 9 meV down to 4 meV on the surface, it is conceivable that some patches of the surface might have undergone a transition to a magnetically ordered phase, similar to the suggestion made by Alexandrov {\it et al.} \cite{Alexandrov&Onur}. If this is the case, one might be able to explain the field-induced hysteresis observed by Nakajima {\it et al.} \cite{Nakajima} in measurements of the magneto-resistance. This scenario would imply that there is a field-induced first-order transition between a polarized paramagnetic surface phase and an antiferromagnetic surface phase caused by the condensation of spin-excitons.

Our calculations show the coupling to spin-excitons does affect the states of the Weyl cone at energies other than just near the resonance energy. In particular, the resonance produces an exponential temperature dependence of the lifetime, renormalized dispersion relation and reduced quasi-particle weight for the near-Fermi-energy surface states. The near Fermi-energy spin-exciton driven scattering is thermally activated but should only appear below the temperature of $T \sim 25$ K at which the bulk spin-exciton seen in neutron scattering experiments becomes intense and sharp \cite{Alekseev3}. The scattering rate has almost completely saturated for temperatures below 10 K.  This temperature is considerably lower than the temperature at which the bulk hybridization gap is first observed \cite{Miyazaki}. Therefore, we argue that the formation of the Fermi-liquid that is responsible for the surface conduction should only occur at very low temperatures and may be responsible for the plateau in the resistivity at 4 K \cite{Nickerson,Allen}.

\section{Acknowledgements}
A.X.G. acknowledges support from the U.S. Army Research Office, under Grant No. W911NF-15-1-0181. S.N. was supported by Forschungszentrum Julich. T.D. acknowledges the NSF IR/D program. Work at Los Alamos National Laboratory was performed under the auspices of the U.S. Department of Energy, Office of Basic Energy Sciences, Division of Materials Science and Engineering.  P. F. S. R. acknowledges a Director's Postdoctoral Fellowship through the LANL LDRD program. T.D. acknowledges the NSF IR/D program. P.S.R. would like to acknowledge support from the US Department of Energy, Office of Basic Energy Sciences, Materials Science through award DE-FG02-01ER45872. The authors would also like to thank S. Borisenko for technical assistance with the measurements and also thank E.D.L. Rienks and O. Rader for numerous discussions and their critical comments.

\end{document}